\documentclass{PoS}
\usepackage{graphicx}
\usepackage{url}
\usepackage{comment}

\title{The Instrument Response Function Format for the Cherenkov Telescope Array}

\ShortTitle{The Instrument Response Function Format for the Cherenkov Telescope Array}

\author{\speaker{John E Ward}$^a$, Javier Rico$^a$ and Tarek Hassan$^a$ for the CTA Consortium\footnote{Full consortium list at http://cta-observatory.org} \\
        \llap{$^a$}Institut de Fisica d'Altes Energies (IFAE), Bellaterra, Spain\\
        E-mail: \email{jward@ifae.es}}

\abstract{The Cherenkov Telescope Array (CTA) is a future ground-based observatory (with two locations, in the Northern and Southern Hemispheres) that will be used in the study of the very-high-energy gamma-ray sky. CTA observations will be proposed by external users or initiated by the observatory, with the resulting measurements being processed by the CTA observatory and the reduced data made accessible to the corresponding proposer. Instrument Response Functions (IRFs) will also be provided to convert the quantities measured by the array(s) into relevant science products (i.e. spectra, sky maps, light curves).

As the response of the telescopes depend on many correlated observational and physical quantities (e.g. gamma-ray arrival direction, energy, telescope orientation, background light, weather conditions etc.) the CTA IRFs could grow into increasingly larger and larger file sizes, which can become unwieldy or impractical for use in specific observation cases. To this end, a customized IRF format (complying with the FITS standard) is under development to reduce the IRF file sizes into more manageable levels.

This proposed format is attractive due to its ability to store multiple parameters (in chosen ranges) relating to instrument performance in both binned and parameterized formats, for various array and observing conditions. Details of the format, preliminary design and testing of the prototype will be provided below.}

\FullConference{The 34th International Cosmic Ray Conference,\\
		30 July- 6 August, 2015\\
		The Hague, The Netherlands}
\begin{document}

\section{Introduction}


CTA\footnote{\url{http://www.cta-observatory.org/}}\cite{CTA_concept} represents the next generation of Imaging Atmospheric Cherenkov Telescopes (IACTs) and consists of two ground-based facilities, one each in the Northern and Southern hemispheres, which will provide full sky coverage in the very-high-energy (VHE) $\gamma$-ray sky from tens of GeV up to approximately one hundred TeV with unprecedented capabilities, improving current experimental sensitivity by over an order of magnitude. In order to have such a broad energy range, each observatory will contain several tens of IACTs of 3 different sizes: 4 Large Sized Telescopes (LSTs), $\sim$ 20 Medium Sized Telescopes (MSTs) and, in the southern-hemisphere site, a large number of Small Sized Telescopes (SSTs).

With CTA, the former model of collaboration-led experiments with private data is being changed towards that of an open observatory, where guest observers may submit observation proposals and have access to the resulting data, along with software analysis tools and support services. The CTA data management \cite{DataMan} must fulfill the requirements of an open observatory and must guarantee reliable processing along with ensuring quality transmission of the data. Furthermore, it must be able to serve data products to a wide and diverse scientific community, made available through community-based standards (e.g. the Virtual Observatory ones). 

One of the sub-components of the CTA "Data Management" project is the Data Model activity \cite{DataModel} which deals with the formats of the data produced by the CTA telescopes from a low (e.g. camera products) to high level (e.g. science products) along with other auxiliary products to enable the analysis \cite{DataPipe} \cite{MC}, description and distribution of data. Two of these auxiliary products for use in generating the final scientific output are the "Lookup Tables" and "Instrument Response Functions", with the latter provided to the guest observers for use in the analysis of their proposed observations. 

Instrument Response Functions (IRFs) are needed by the user to translate the collective reconstructed variables (particle nature (ID'), direction (p'), and Energy (E')) from all detected events into to a particle flux (via de-convolution of the instrument response through a combination of estimator Probability Density Functions (PDFs)).

For IACTs arrays, storing LUTs and providing IRFs to the community is very challenging due to the large observational parameter space, the need to provide sufficient Monte-Carlo statistics for a broad energy range, the various possible array/detector layouts, the long- and short-term variation in performance due to atmospheric effects, as well as detector performance with age.  All of these need to be considered, as well as the potential for very large file sizes, when designing and implementing a new storage format. 

\section{Task Overview}

At several steps in the data reduction/processing procedure, the data from a given observation needs to be combined with those representing the response of the instrument (or a given part of it). Those instrument-related data are to be described and stored in an Instrument Response Model (IRM) format. 

The IRM is divided into two parts; the stored mid-level data used to transform the reconstruction pipeline data into the accessible estimated physical quantities (such as energy reconstruction, particle arrival direction and particle ID) are generally referred to as Lookup Tables (LUTs) while the response functions needed to generate the final scientific products (e.g. particle flux) are termed High-Level Instrument Response Functions (HLIRFs). In addition, the HLIRFs are an essential input to compute the feasibility and expected performance of new scientific and/or technical CTA projects. 

Due to the large parameter space of potential CTA observations and reconstructed variables, the IRM format needs to provide sufficient flexibility to handle large volumes of information while also complying with the FITS format \cite{FITS_format} standard (a requirement of the CTA software project). The inputs to the IRM format and how the information is stored also needs to be defined. 

\begin{figure}
\centerline{\includegraphics[width=0.90\textwidth]{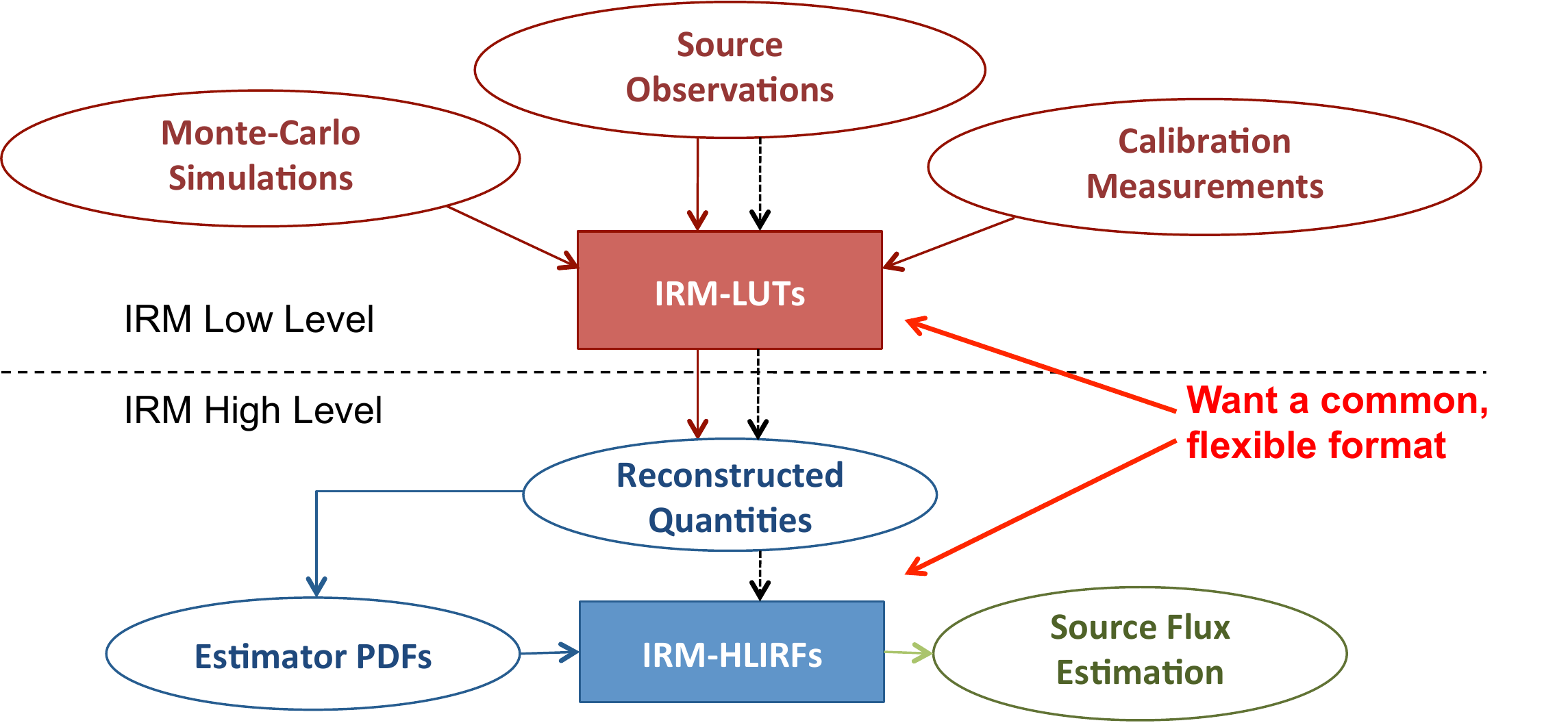}}
\caption{Flow chart representing the various processes that will interface with the IRM format}
\label{fig:IRM_image}
\end{figure}

We have identified several parts composing or directly interacting with the IRM and its task (see Figure \ref{fig:IRM_image}). Overall, the IRM is constructed with data obtained from simulations, dedicated measurements and results from the calibration and reconstruction pipelines (e.g. for example, samples of full Monte-Carlo simulated gamma-ray showers for the computation of the effective collection area of a given sub-array). In general, these data correspond to a category different from the IRM (e.g. to Monte Carlo and raw processed data (DL1,\cite{DataMan})). However, they are an important element to the IRM data format, since they must be produced/acquired/provided in sufficient amount and variety to allow for a full construction of the IRM in all the relevant observational conditions, and also because they must be referred to or linked by the IRM metadata.

At the first level within the IRM, the LUTs contain the input data for computing the value of one or several reconstructed physical quantities, starting from the values of one or several measured quantities (or reconstructed in an earlier stage of the data reduction process). In addition to the input data, the output produced by the LUTs may be influenced by one or several configuration parameters. The LUTs are produced by the Data Pipeline and require a reference to the particular version used for a given data set, contained within their IRM meta-data. Therefore, the group responsible for the IRM must work in close collaboration with those responsible for the Data Pipeline.

The higher-level state of the IRM utilizes the LUTs to provide the relevant estimates of the physical properties of the measured data for use in the effective area, PSF and energy dispersion PDFs that will make up the HLIRFs for use in particle flux estimation.

We have identified the following processes as needing input from the IRM:

\textbf{LUTs}

- Shower geometrical reconstruction, i.e. the computation of the shower geometrical parameters (such as incident direction, core impact point, shower maximum and others) starting from the reconstructed images for the different involved cameras.

- Shower calorimetric reconstruction, i.e. the computation of the shower reconstructed energy starting from the shower geometrical parameters plus the image parameters for the different cameras.

- Particle nature reconstruction, i.e. the determination of the particle nature (e.g. gamma-ray or cosmic ray) starting from the shower calorimetric and geometrical parameters plus the image parameters for the different cameras.

\textbf{HLIRFs}

- Flux reconstruction, i.e. the determination of the number of signal events (particle per area, time, unit energy and solid angle) from a given region of the observed field of view. This is needed to determine, e.g. the spectrum and light curve from a given gamma-ray source, or the gamma-ray sky-map in a given field of view.

Another important issue regarding the IRM is that files must be produced for all relevant observational conditions. The parameters defining such conditions must necessarily be part of the IRM format so that a link between a given data set and the corresponding IRM file can be established, during data reduction/analysis, but also for posterior reference and reproducibility. In general, IRM files shall be constructed in bins of the different relevant observational parameters, in such a way that the provided IRM file can be considered a good description of the instrument within the whole bin and they can achieve the sensitivity and systematic errors requirements of the CTA observatory. This is especially relevant for those cases for which the construction of IRM files may need a large deal of resources (e.g. production of relatively large Monte-Carlo simulated samples for effective collection area determination). We have identified the following list of observational parameters that must be taken into account when constructing and applying the IRMs:

\begin{itemize}
  \item Array configuration, i.e. which telescopes are involved in a given observation 
  \item Array observation mode, i.e. what are the relative telescope orientations and tracking modes. This may include the following modes: pointed observations, divergent pointing, drift scans 
  \item Individual telescope orientations
  \item Atmospheric conditions
  \item Light illumination, i.e. by NSB, moon and/or others which will lead to different trigger settings.
  \item Hardware conditions, including those which are configurable and those which will naturally change with time, e.g. telescope optical performance, PMT gains, trigger levels and logics, failing components, and others.
  \item Analysis methods and parameters: algorithms, cuts, etc.
\end{itemize}

\newpage
\section{General Requirements}
\label{gen_req}

The requirements of the IRM format have been defined such that the format must:

\begin{itemize}
\item contain enough information so that together with the reconstruction methods from the Data Pipeline it can be used to compute all the relevant reconstructed variables in the data processing chain.
\item be reproducible (i.e. include the information of the data and of the methods that were used to build them, also must include the information of the methods that they are to be used with).
\item be provided for all relevant observational conditions, including the information of the observational conditions for which they are applicable.
\item include all relevant configurable parameters (e.g. analysis cuts) used by the relevant Data Pipeline methods used during their construction.
\item contain the necessary meta-data to provide a reference on how the Pipelines analysis converted calibrated pixel information into individual telescope image information (image reconstruction).
\item contain the necessary meta-data to provide information on how the shower geometrical parameters, such as incident direction, ground impact point and shower maximum altitude were calculated from the individual telescope images (shower geometrical reconstruction). 
\item allow estimation of the shower energy with the best possible precision using the reconstructed image and shower geometrical parameters (energy estimation).
\item allow estimation of the Flux (units of cm$^{-2}$ s$^{-1}$) as a function of the energy, incident direction and time (flux estimation)

\end{itemize}

\section{IRM Format Scheme}

\begin{figure}
\centerline{\includegraphics[width=0.9\textwidth]{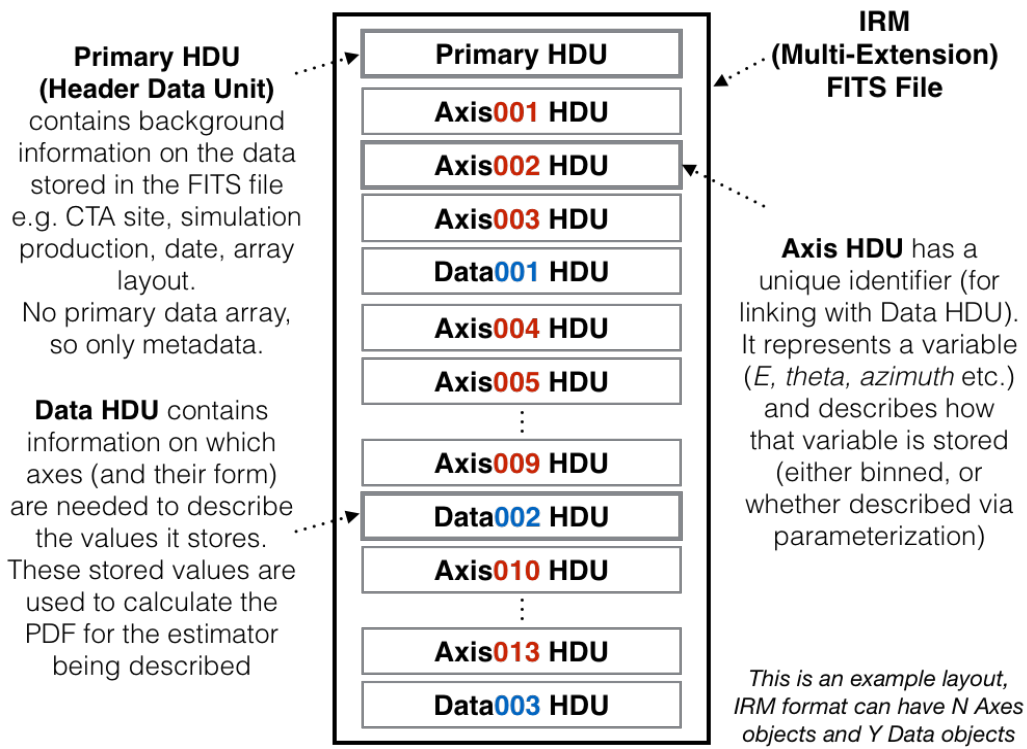}}
\caption{IRM file diagram. HDU stands for Header Data Unit.}
\label{fig:IRF_layout}
\end{figure}

The IRM format currently in development to fulfill the requirements stated in Section \ref{gen_req} is shown in a diagrammatic form in Figure \ref{fig:IRF_layout}. The format consists of a multi-extension N-dimensional FITS file that contains several Header Data Units (HDUs). This file can either be a "global" (considered like a database) file containing all the relevant information needed to construct a LUT or HLIRF for any observation in CTA, or can indeed be built using a much smaller parameter range that may only be needed for a small-scale observation. Whether used in the global or local case, the format of the IRM FITS file stays the same. Therefore, this format strategy will allow smaller, user/observation-specific HLIRFs to be extracted from a larger database file and provided to the analyzer in a much more manageable size. 

A further memory-saving strategy can be implemented whereby several variables and PDFs can be parameterized, with the resulting parameter values stored instead. However, the IRM format is suitably flexible to allow either/or strategy to be implemented in CTA as the need arises. 

\subsection{Primary HDU}

The "primary" HDU contains meta-data on the information that is being stored by the file, for instance: CTA site, simulation version, array layout etc. There is no data stored in the primary HDU array. Additional meta-data can be added as needed to the primary header. Two other HDUs are utilized in the file to provide the relevant information to the end user.

\subsection{Axis HDU}

The Axis HDU represents a variable (e.g. Energy, offset angle, azimuth etc.) and describes if that variable is stored by binning or a parameterization (see Figure \ref{fig:IRF_layout_2}). The header of the Axis HDU contains meta-data on the length of the axis, the type, the variable and/or a definition of validity for the range of the parameterization. The data array of the Axis HDU contains the lower bin limits (in the case of a binned axis), or information relating to the form of the function used for the parameterization. The axis variable and type is supplied by FITS keywords which can be added or subtracted by the developer. Currently, the allowed axis variables are NoVarType, Energy, Energy (true), Energy (reconstructed), theta (for offset angle), phi (for azimuthal angle), ID (particle ID) and VarMax (defining the maximum range of validity for any parameterization). 

\subsection{Data HDU}

\begin{figure}
\centerline{\includegraphics[width=0.9\textwidth]{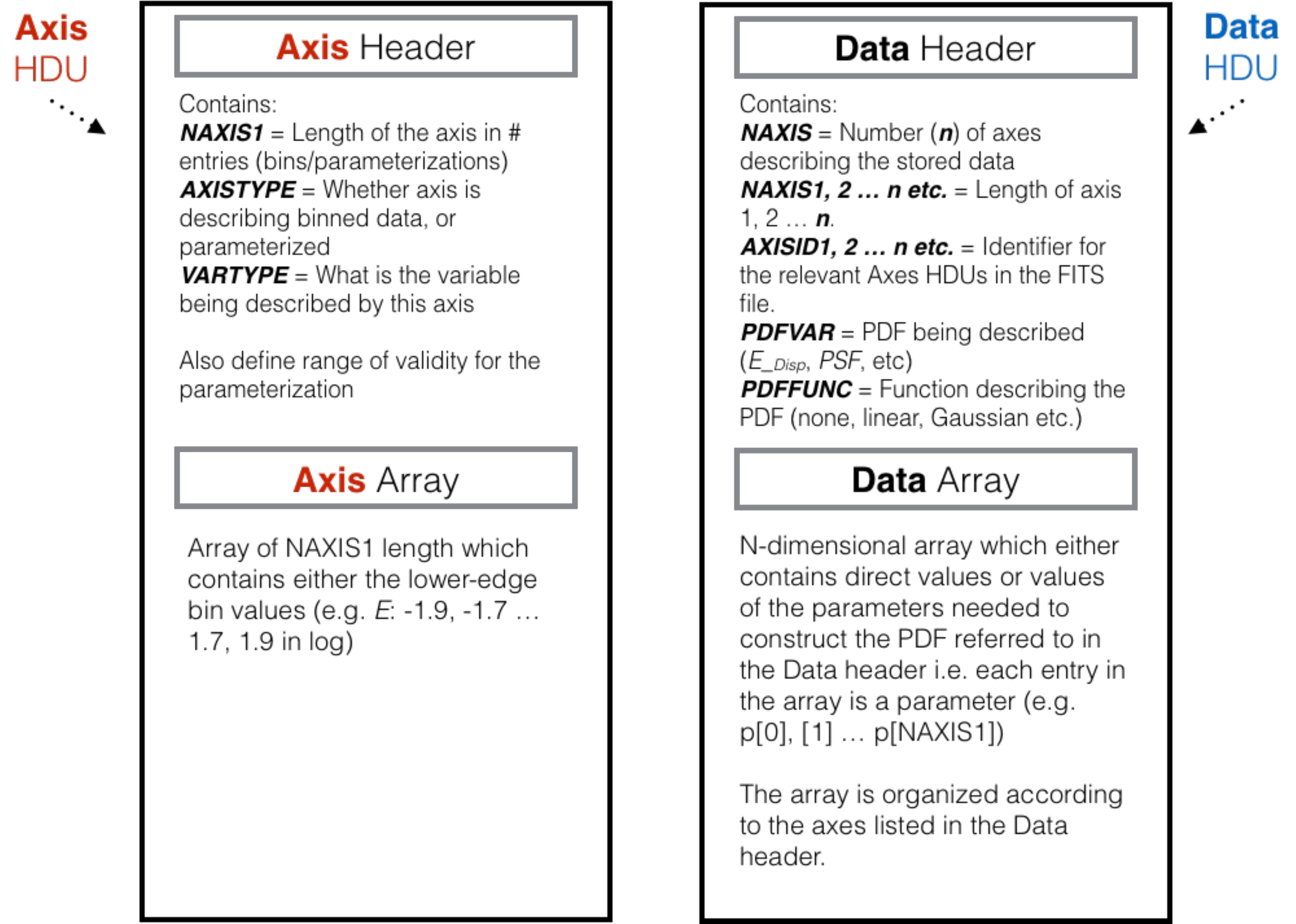}}
\caption{IRM Axis and Data HDU breakdown}
\label{fig:IRF_layout_2}
\end{figure}

The Data HDUs (see see Figure \ref{fig:IRF_layout_2}) contain meta-data (in the header) that explain which Axis objects are needed to describe the information contained in the data array itself along with the PDF being stored, and which function has been used (if any) to characterize it. This stored information is used to construct the PDFs in the case of HLIRFs. The Data HDU can contain a N-dimensional array with the corresponding N axis objects needed to rationalize it fully. The PDFs are described by FITS keywords, with the current implementation allowing no PDF (i.e. direct values, not a probability) or PDFs describing efficiency, Energy dispersion, point spread function, background rate, background rate per squared degree, differential sensitivity, effective area and effective area with no arrival-direction cut (i.e. no theta$^2$ cut).

\section{IRM Status}

Currently, the IRM format has been written and implemented in C/C++ utilizing the CFITSIO library. A tool to convert ROOT histograms of CTA performance files into the IRM format has been written and tested. Studies on the merging and rebuilding of IRM data-cubes, and the ability to add additional information and axes to database cubes \emph{a posteriori} have also been undertaken. The IRM format can also supply full Migration Matrices or parameterized versions as needs be. 

The next step currently being undertaken is that detailed discussions on the best way to utilize the IRM format in CTA science tools packages (Ctools, PyFACT etc.), with an aim to be ready to use the format with the first constructed prototype telescopes of CTA. Development work to this effect has already begun.

\newpage
\acknowledgments{
We gratefully acknowledge support from the agencies and organizations 
listed under Funding Agencies at this website: http://www.cta-observatory.org/.
This work is partially funded by the ERDF under the Spanish MINECO grant FPA2012-39502.
}

\end{document}